\documentclass[11pt]{article}
\usepackage{epsfig}

\begin{document}

\newcommand{\bm}[1]{\mbox{\boldmath$#1$}}
\def\mvec#1{{\bm{#1}}}   

\title{Asymmetric Uncertainties: \\
Sources, Treatment  and Potential Dangers}
\author{G.~D'Agostini \\
Universit\`a ``La Sapienza'' and INFN, Roma, Italia \\
(giulio.dagostini@roma1.infn.it, www.roma1.infn.it/\,$\tilde{ }$\,dagos)
}
\date{}

\maketitle
\begin{abstract}
The issue of asymmetric uncertainties resulting from
fits,  nonlinear propagation and  
systematic effects is
reviewed. It is shown that, in all cases, 
whenever a published result is given with asymmetric uncertainties,
the {\it value} of the physical quantity of interest
{\it  is biased} with respect to what would be obtained using at best
all experimental and theoretical information that contribute 
to evaluate the combined  uncertainty. 
The probabilistic solution to the problem is provided
both in exact and in approximated forms. 
\end{abstract}

\section{Introduction}
We often see published results in the 
form
$$\mbox{`best value'}\ ^{+\Delta_+}_{-\Delta_-}\,,$$
where $\Delta_+$ and $\Delta_-$ are {\it usually} 
positive.\footnote{For examples of measurements 
having $\Delta_+$ and $\Delta_-$ with all combinations of  
signs, see public online tables of Deep Inelastic 
Scattering results.\cite{HERA}
I want to make clear since the very beginning
that it is not my intention 
to blame experimental or theoretical teams which have
reported in the past asymmetric uncertainty, 
because we are all victims
of a bad tradition in data analysis. At least,
when asymmetric uncertainties have been given, 
 there is some chance to correct the result, as described in 
Sec.~\ref{sec:thumb}. Since some asymmetric contributions to the 
global uncertainties almost unavoidably happen  in complex experiments,
I am more worried of collaborations that never
arrive to final asymmetric
uncertainties, because I must imagine they have 
symmetrised somehow the result but, I am afraid, without 
applying the proper shifts to the `best value' to take into account
asymmetric contributions, as it will be discussed in the present paper.} 
As firstly pointed out in Ref.~\cite{ConMirko} and discussed in a simpler
but more comprehensive way in Ref.~\cite{BR}, this practice 
is far from being acceptable and, indeed, could bias the 
believed value of important physics quantities. 
The purpose of the present paper is, summarizing and somewhat 
completing the work done
in the above references, 
to remind where asymmetric uncertainty stem from 
and to show why, as they are usually treated,
they bias the value of physical quantities,
either in the published result itself or in subsequent analyses. 
Once the problems are spotted, the remedy is straightforward,
at least within the Bayesian framework (see e.g. 
\cite{BR}, or \cite{RPP_GdA} and \cite{RPP_Dose} for 
recent reviews). In fact the Bayesian approach is 
conceptually based 
on the intuitive idea of probability, and formally grounded on 
the basic rules of probability (what are usually known as the
probability `axioms' and the `conditional probability definition')
plus logic.
Within this framework 
many methods of `conventional' statistics are reobtained,
 as approximations of general solutions,
under well stated conditions of validity.
Instead, in the conventional, frequentistic approach
{\it ad hoc} formulae, prescriptions and un-needed principles
are used, often without understanding what is behind
these methods -- before a `principle' there is nothing! 

The proposed Bayesian solutions to cure the troubles 
produced by the usual treatment of
asymmetric uncertainties
is to step up from approximated methods to the
more general ones (see e.g. Ref.~\cite{BR}, 
in particular the top down approximation 
diagram of Fig.~2.2). 
In this paper we shall see, for example,
how $\chi^2$ and minus log-likelihood fit `rules'  
can be derived from the Bayesian inference formulae
as approximated methods and what to
do when the underlying conditions do not hold. 
We shall encounter a similar situation regarding
standard formulae to propagate uncertainty.

Some of the issues addressed here and in 
 Refs. \cite{ConMirko} and \cite{BR}
have been recently 
brought to our attention by Roger Barlow~\cite{Barlow},
who proposes frequentistic ways out. 
Michael Schmelling had also addressed questions related to
`asymmetric errors', particularly related to 
the issue of weighted averages~\cite{Schmelling}.
The reader is encouraged to
read also these references to form his/her idea about
the spotted problems and the proposed solutions. 

In Sec. \ref{sec:propagation} the issue of propagation of
uncertainty is briefly reviewed at an elementary level 
(just focusing on the sum of
uncertain independent variables -- i.e. 
no correlations considered) though taking into account 
asymmetry in probability density functions (p.d.f.) of the
{\it input} quantities. In this way we understand 
what `might have been done' 
(we are rarely in the positions to exactly know ``what has been done'')
by the authors who publish 
asymmetric results and what is the danger of improper use of  
such a published `best value' -- {\it as is} -- in subsequent analyses. 
Then, Sec.~\ref{sec:sources} we shall see in  
where asymmetric uncertainties stem from and what to do in order
to overcome their potential 
troubles. This will be done in an exact way and, whenever is possible,
in an approximated way. 
Some rules of thumb to roughly recover  sensible probabilistic
quantities (expected value and standard deviation) from 
results published with asymmetric uncertainties 
will be given in Sec.~\ref{sec:thumb}. 
Finally, some conclusions will be drawn.

\section{Propagating uncertainty}\label{sec:propagation}
Determining the value of a physics quantity is seldom
an end in itself.
In most cases the result is used, together with  other 
experimental and theoretical quantities, to 
calculate the value of other quantities of interest. 
As it is well understood, uncertainty on the value of each 
ingredient is propagated into uncertainty on the 
final result. 

If uncertainty is quantified by probability, as it is commonly done 
explicitly or implicitly\footnote{Perhaps the reader would
be surprised to learn that in the conventional statistical
approach there is no room for probabilistic 
statements  about the value of physics quantities
(e.g. 
``the top mass is between 170 and 180 GeV with 
such percent probability'', or ``there is 95\% probability that the
Higgs mass is lighter than 200 GeV''),
calibration constants, and so on, as discussed extensively
in Ref.~\cite{Maxent98}.}
 in physics, the propagation
of uncertainty is performed using rules based on probability theory.
If we indicate by $\mvec X$ the set 
(`vector') of input quantities and by $Y$ the 
final quantity, given by the function $Y=Y(\mvec X)$ of the
input quantities, the most general propagation formula
(see e.g. \cite{BR})
 is given  by (we stick to continuous variables): 
\begin{equation}
f(y) = \int\! \delta[y-Y(\mvec x)]\cdot f(\mvec x)\, \mbox{d}\mvec x\,,
\label{eq:prop_general}
\end{equation} 
where $f(y)$ is the p.d.f. of $Y$,
$f(\mvec x)$ stands for the joint p.d.f. of $\mvec X$ 
and $\delta$ is the Dirac delta
(note the use of capital letters to name variables and small
letters to indicate the values that variables may assume).  
The exact evaluation of Eq.~(\ref{eq:prop_general}) is often challenging,
but, as discussed in Ref.~\cite{BR}, this formula has a 
nice simple interpretation
that makes its Monte Carlo implementation conceptually easy.

As it is also well known, often there is no need to go through the 
analytic, numerical or Monte Carlo evaluation of Eq.(\ref{eq:prop_general}),
since linearization of  $Y(\mvec x)$ around the expected value  
of $\mvec X$ (E[$\mvec X$]) makes the calculation of 
expected value and variance of $Y$ very easy, using the well known 
standard propagation formulae, that for uncorrelated input quantities are
\begin{eqnarray}
\mbox{E}[Y] & \approx & Y(\mbox{E}[\mvec X]) 
\label{eq:prop_approx_E} \\
\sigma^2(Y) & \approx &
\sum_i \left(\left.\frac{\partial Y}{\partial X_i}
       \right|_{\mbox{E}[\mvec X]}\right)^2\, \sigma^2(X_i)\,.
\label{eq:prop_approx_sigma}
\end{eqnarray} 
As far as the shape of $f(y)$, a Gaussian one is usually assumed, 
as a result of the central limit theorem. 
Holding this assumptions,  
$\mbox{E}[Y]$ and $\sigma(Y)$ is all what we need. 
$\mbox{E}[Y]$ gives the `best value', and 
probability intervals, upper/lower limits and so on 
can be easily calculated. 
In particular, within the Gaussian approximation,
the most believable value ({\it mode}), the barycenter of the 
p.d.f. ({\it expected value}) and the value that separates
two adjacent 50\% probability intervals ({\it median}) coincide.
If $f(y)$ is asymmetric this is not any longer true and one 
needs then to clarify what `best value' means,
which could be one of the above three {\it position parameters},
or something else (in the Bayesian approach
`best value' stands for expected value, unless differently specified).

\begin{figure}[t]
\begin{center}
\begin{tabular}{|cccc|}\hline
&&& \\
& \multicolumn{1}{l}{\small $\begin{array}{lcl} \mbox{E}(X) &=& 0.17 \\
 \sigma(X) & = & 0.42 \\
\mbox{mode} &=& 0.5 \\
\mbox{median} &=& 0.23 
\end{array}$} & & 
\multicolumn{1}{c|}{\small $\begin{array}{lcl} \mbox{E}(Y) &=& 0.34 \\
 \sigma(Y) & = & 0.59 \\
\mbox{mode} &=& 0.45  \\
\mbox{median} &=& 0.37 
\end{array}$} \\
&&& \\
\hspace{0.3cm}{\Large $2\times$} & 
\epsfig{file=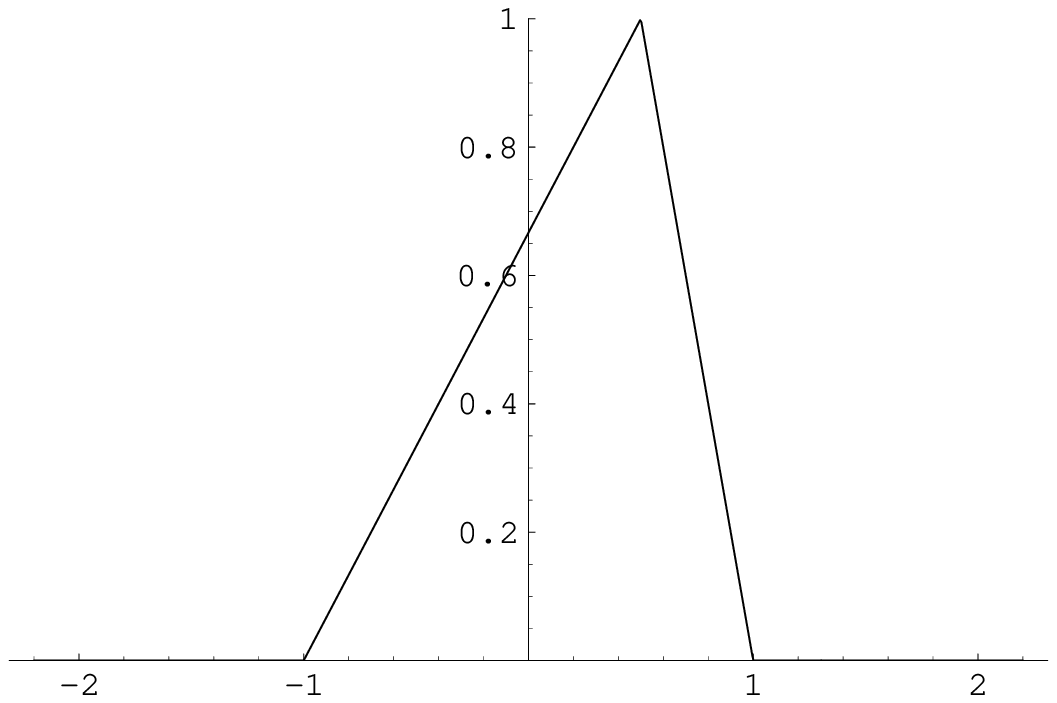,width=0.25\linewidth,
height=4.22cm,clip=} 
&{\Large $\Longrightarrow$} & 
\epsfig{file=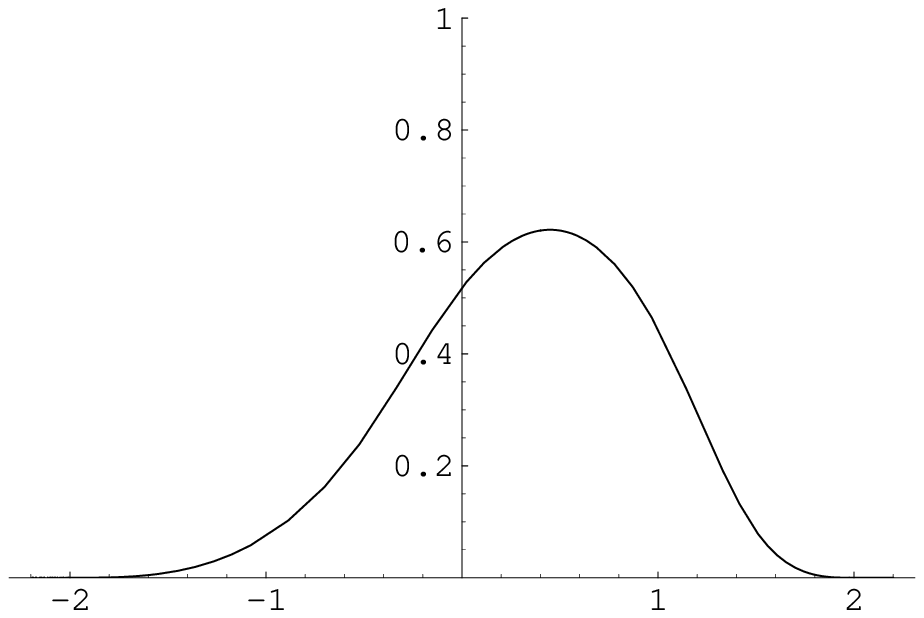,width=0.46\linewidth,height=4.1cm,clip=} \\
\hline
\end{tabular}
\end{center}
\caption{\small 
Distribution of the
sum of two independent quantities, each described by an asymmetric
 triangular p.d.f. self-defined in the left plot.
The resulting p.d.f.
(right plot) has been calculated analytically making use of
Eq.(\ref{eq:prop_general}). 
This figure corresponds to Fig.~4.3 of Ref.~\cite{BR}.}
\label{fig:2triang}
\end{figure}

Anyhow, Gaussian approximation is not the main issue here and, in most 
real applications, characterized by several contributions to the 
combined uncertainty about $Y$,   
this approximation is a reasonable one, even when some of the input
quantities individually contribute asymmetrically. 
My concerns in this paper  are more related to the evaluation
of $\mbox{E}[Y]$ and $\sigma(Y)$ when
\begin{enumerate}
\item
instead of 
Eqs.~(\ref{eq:prop_approx_E})--(\ref{eq:prop_approx_sigma}),
 {\it ad hoc} 
propagation prescriptions are used 
in presence of asymmetric uncertainties;
\item
linearization implicit in 
Eqs.~(\ref{eq:prop_approx_E})--(\ref{eq:prop_approx_sigma})
is not a good approximation.
\end{enumerate}
Let us start with the first point, considering, as
an easy academic example, input quantities described by the 
 asymmetric triangular distribution
shown in the left plot of Fig.~\ref{fig:2triang}. 
The value of $X$ can range between $-1$ and $1$, with a 
`best value', in the sense of maximum probability value,
of 0.5. The interval $[-0.16, +0.72]$ gives a 68.3\% probability
interval, and the `result' could be reported
as $X_1=0.50^{+0.22}_{-0.66}$. This is not a problem as long as we known
what this notation means and, possibly, know the shape of $f(x)$. 
 The problem arises when we want 
to make use of this result and we do not have access to 
$f(x)$ (as it is often the case), or we make improper use
of the information [i.e. in the case we are aware of $f(x)$]. 
Let us assume, for simplicity,
to have a second independent quantity, $X_2$, described exactly by
the same p.d.f. and reported in the same way: 
$X_2=0.50^{+0.22}_{-0.66}$. Imagine we are now interested
to the quantity $Y=X_1+X_2$. How to report the result about $Y$, based
on the results about $Y_1$ and $Y_2$? Here are some common,
but {\it wrong} ways to give the result:
\begin{itemize}
\item
{\it asymmetric uncertainties added in quadrature}: 
$Y=1.00^{+0.31}_{-0.93}$;
\item
{\it asymmetric uncertainties added linearly}: 
$Y=1.00^{+0.44}_{-1.31}$.
\end{itemize} 
Indeed, in this simple case we can calculate 
the integral (\ref{eq:prop_general}) 
 analytically, obtaining the curve shown in the 
plot on the right side of 
Fig.~\ref{fig:2triang}, where several position and
shape parameters have also been reported. The `best value' of $Y$,
meant as expected value (i.e. the barycenter of the p.d.f.)
comes out to be 0.34. Even those who like to think at the
`best value' as the value of maximum probability (density) 
would choose 0.45 (note that in this particular example the mode of the sum
is smaller than the mode of each addend!). Instead, 
a `best value' of $Y$ of 1.00 obtained by the {\it ad hoc}
rules, unfortunately often used in physics, corresponds neither
to mode, nor to expected value or median. 

The situation would have been much better if
expected value and standard deviation of $X_1$ and $X_2$
had been reported (respectively 0.17 and 0.42). Indeed, these 
are the quantities that
matter in `error propagation', because the {\it theorems upon which
propagation formulae rely} --- exactly in the case $Y$ is a linear combination
of $\mvec X$, or approximately in the case linearization has been performed ---
{\it speak of expected values and variances}. 
It is easy to verify from the numbers in Fig.~\ref{fig:2triang} 
that exactly the correct values of $\mbox{E}[Y] = 0.34$ and 
$\sigma(Y)=0.59$ would have been obtained.
Moreover, one can see that 
expected value, mode and median of $f(y)$ do not differ much from
each other, and the shape of $f(y)$ resembles a somewhat
skewed Gaussian. When $Y$ will be combined with other quantities
in a next analysis its slightly non-Gaussian shape 
will not matter any longer. Note that we have achieved this nice
result already with only two input quantities. If we had a few 
more,
already $Y$ would have been much Gaussian-like. Instead, 
performing a bad combination of several quantities all skewed in the
same side would yield `divergent' 
results\footnote{The reader might be curious to know what would happen
in case of bad combinations of input 
quantities with skewness of mixed signs.
Clearly there will be some compensation that lowers the risk of
strong bias. As an academic exercise, let think of five independent 
variables each
described by the triangular distribution of Fig.~\ref{fig:2triang}
and five others each described by a p.d.f. which is its mirror reflexed
around $x=0.5$ 
[$0\le X\le 2$, $\mbox{mode}(X)=0.5$,  $\mbox{E}[X] = 0.83$ and 
$\sigma(X)=0.42$]. The correct combination of the ten variables
gives $Y=5.00\pm 1.33$, while
adding the modes and combining quadratically left and right deviations
we would get $5.00\pm  1.54$.}:
for $n=10$ we get, 
using a quadratic combination of left and right deviations,
$Y=5.00^{+0.69}_{-2.07}$ versus the correct $Y=1.70\pm 1.32$. 

As conclusion from this section I would like to make some points:
\begin{itemize}
\item
in case of asymmetric uncertainty on a quantity,
it should be avoided to report {\it only} 
most probable value and a probability interval 
(be it 68.3\%, 95\%, or what else);
\item
expected value, meant as barycenter of the distribution, 
as well as standard deviations should {\it always} be reported,
providing also the shape of the distribution (or its summary
in terms of shape parameters, or even a parameterization of the 
log-likelihood function in a 
polynomial form, as done e.g. in Ref.~\cite{ZEUS}), 
 if the distribution is asymmetric or non trivial.
\end{itemize}
Note that the propagation example shown here is the most
elementary possible. The situation gets more complicate if
also nonlinear propagation is involved (see Sec.~\ref{ss:nonlinear})
or when quantities are used in fits (see e.g. Sec.~12.1 of
 Ref.~\cite{BR}). 

Hoping that the reader is, at this point, at least
worried about the effects of badly treated asymmetric uncertainties, 
let us now review
the sources of asymmetric uncertainties.

\section{Sources of asymmetric uncertainties}\label{sec:sources}
\subsection{Non parabolic $\chi^2$ or log-likelihood curves}
The standard methods in physics to adjust theoretical parameters to
experimental data are based on {\it maximum likelihood principle} 
ideas.
In practice, depending on the situation, the `minus log-likelihood'
of the parameters 
[$\varphi(\mvec\theta;\mbox{data})=-\ln L(\mvec\theta;\mbox{data})$] 
or the $\chi^2$ function of the parameters [i.e. the function
$\chi^2(\mvec\theta;\mbox{data})$] are minimized. 
The notation used reminds
that $\varphi$ and $\chi^2$ are seen as 
mathematical function of the parameters 
$\mvec\theta$, with the data acting as `parameters' of the functions. 
As it is well understood,  a part from an irrelevant constant non 
depending on fit parameters,
$\varphi$ and $\chi^2$ differ by just a factor of two
when the likelihood, seen as a joint probability function or a p.d.f.
of the data,
 is a (multivariate) Gaussian distribution of the data:
 $\varphi=\chi^2/2+k$ (the constant $k$ is often neglected,
since we concentrate on the terms which depend on the fit 
parameters -- but sometimes $\chi^2$ and minus log-likelihood
might differ by terms depending on fit parameters!).
For sake of simplicity, let us take one parameter fit. 
Following the  usual practice, we indicate  the parameter by $\theta$
(though this fit parameter is just any of the input quantities
$\mvec X$ of Sec.~\ref{sec:propagation}). 

If  $\varphi(\theta)$ or  $\chi^2(\theta)$ have a nice parabolic 
shape, 
the likelihood is, apart a multiplicative factor, 
a  Gaussian function\footnote{But not yet a probability function! 
The likelihood has the probabilistic meaning of a joined 
p.d.f. of the data given $\theta$, 
and not the other way around.} of $\theta$.  
In fact, as is well known from calculus,
any function can be approximated to a parabola in the vicinity
of its minimum. 
Let us see in detail the expansion of 
 $\varphi(\theta)$ around its minimum 
$\theta_m$:
\begin{eqnarray}
\varphi(\theta) &\approx& \varphi(\theta_m) + 
\left.\frac{\partial \varphi}{\partial \theta}\right|_{\theta_m}
\!\!(\theta-\theta_m) + 
\left.\frac{\partial^2 \varphi}{\partial \theta^2}\right|_{\theta_m}
\!\!\frac{1}{2}\,(\theta-\theta_m)^2 \\
&\approx& \varphi(\theta_m) + 
\frac{1}{2}\,\frac{1}{\alpha^2}\,(\theta-\theta_m)^2\,, 
\end{eqnarray}
where the second term of the r.h.s. vanishes by definition 
of minimum and we have indicated with $\alpha$ the 
inverse of the second derivative at the minimum. 
Going back to the likelihood, we get:
\begin{eqnarray}
L(\theta; \mbox{data}) 
&\approx & \exp{\left[- \varphi(\theta_m)\right]} \cdot
 \exp{\left[-\frac{1}{2}\,\frac{1}{\alpha^2}\,(\theta-\theta_m)^2\right]} \\
 &\approx & k\,  \exp{\left[-\frac{(\theta-\theta_m)^2}
                                               {2\,\alpha^2}\right]}\,,
\end{eqnarray}
apart a multiplicative factor, this is  
`Gaussian'
centered in
 $\theta_m$ with standard deviation 
$(\partial^2 \varphi/\partial \theta^2|_{\theta_m})^{-2}$.
However, although this function is mathematically
a Gaussian, it does not have yet the meaning of 
probability density $f(\theta\,|\,\mbox{data})$ in an inferential 
sense, i.e. describing our knowledge about $\theta$ in the light 
of the experimental data. In order to do this, we need to process
the likelihood through Bayes theorem, 
which allows {\it probabilistic inversions} to be achieved using
basic rules of  probability theory and logic. 
Besides a conceptually irrelevant normalization factor
(that has to be calculated at some moment) the Bayes formula is 
\begin{eqnarray}
f(\theta\,|\,\mbox{data}) &\propto& f(\mbox{data}\,|\,\theta)
       \cdot f_0(\theta)\,.
\label{eq:Bayes}
\end{eqnarray}
We can speak now about the ``probability that $\theta$ is 
within a given interval'' and calculate it, together
with expectation of $\theta$, standard deviation 
and so
on.\footnote{$\theta$ has not a probabilistic 
interpretation in the frequentistic approach,
and therefore we cannot
 speak consistently, in that framework, 
about its probability, or determine expectation,
standard deviation and so on. Most physicists do not even know
of this problem and think these are irrelevant semantic quibbles.
However, it is exactly this contradiction between intuitive thinking and 
cultural background\cite{Maxent98}  that causes wrong 
scientific conclusions, like those discussed in this paper.
} 
If the {\it prior}  $f_0(\theta)$ is much 
vaguer that what the 
data can teach us (via the likelihood),  
then it can be re-absorbed in the normalization constant,
and we get:
\begin{eqnarray}
f(\theta\,|\,\mbox{data}) &\propto& f(\mbox{data}\,|\,\theta) 
                         = L(\theta;\mbox{data}) 
\label{eq:final_gen} \\
\mbox{i.e}\ \  \ \ \ && \nonumber\\
   &\propto& \exp{\left[-\varphi(\theta;\mbox{data})\right]} \label{eq:final_phi} \\
\mbox{or}\ \  && \nonumber\\
   &\propto& \exp{\left[-\frac{\chi^2(\theta;\mbox{data})}{2}\right]}
\label{eq:final_chi2}  \\
 \mbox{ parabolic} \ \varphi\ \mbox{or}\ \chi^2\,: \ 
 && \nonumber \\
\rightarrow\  f(\theta\,|\,\mbox{data}) & = & 
\frac{1}{\sqrt{2\pi}\, \sigma_\theta}\,
   \exp{\left[-\frac{(\theta-\mbox{\small E}[\theta])^2}
                    {2\,\sigma_\theta^2}\right]}\,.   
\end{eqnarray}
If this is the case, it is a simple exercise to show that
\begin{enumerate}
\item[{\it a})]
$\mbox{E}[\theta]$ is equal to $\theta_m$ which minimizes the 
$\chi^2$ or $\varphi$. 
\item[{\it b})]
$\sigma_\theta$ can be obtained by the famous conditions
$\Delta \chi^2 = 1$ or $\Delta \varphi = 1/2$, respectively,
or by the second derivative around $\theta_m$:
$\sigma_\theta^{-2} = 
1/2\times \left.(\partial^2 \chi^2/\partial \theta^2)
          \right|_{\theta_m}$
or 
$\sigma_\theta^{-2} = 
\left.(\partial^2 \varphi/\partial \theta^2)
          \right|_{\theta_m}$,
respectively.
\end{enumerate}
Though in the frequentistic approach language and methods are usually
more convoluted (even when the same numerical results of the Bayesian
reasoning are obtained), due to the fact that probabilistic statements
about physics quantities and fit parameters are not
allowed in that approach, it is usually accepted that 
the above rules $a$ and $b$ are based on the parabolic behavior of 
the minimized functions. 
When this approximation does not hold, the 
frequentist has to replace a prescription by other prescriptions
that can handle the exception.\footnote{It is a matter of fact that
the habit in the particle physics community of applying uncritically 
the $\Delta \chi^2=1$ or $\Delta \varphi=1/2$ is related 
to the use of the software package {\it MINUIT}\cite{MINUIT}.
Indeed, {\it MINUIT} can calculate the parameter variances
also from the $\chi^2$ or $\varphi$ curvature at the minimum 
(that relies on the same hypothesis
upon which  the  $\Delta \chi^2=1$ or $\Delta \varphi=1/2$
rules are based). But when the  $\chi^2$ or $\varphi$ 
are no longer parabolic, the standard deviation calculated from the
curvature  
 differs from that  
of the  $\Delta \chi^2=1$ or $\Delta \varphi=1/2$
(in particular, when the minimized function is asymmetric 
the latter rules give two values, the (in-)famous $\Delta_\pm$
we are dealing with).
 People realize 
that the curvature at the minimum depends from the local behavior 
of the minimized curve, and the   $\Delta \chi^2=1$ or $\Delta \varphi=1/2$
rule is typically more stable. Therefore, in particle physics 
the latter rule has become {\it de facto} a standard to evaluate
`confidence intervals' at different `levels of confidence'
(depending of the value of the  $\Delta \chi^2$ or $\Delta \varphi$). 
But, unfortunately, when those famous curves are not parabolic, 
numbers obtained by these rules might loose completely a probabilistic meaning.
[Sorry, a frequentist would object that, indeed, these numbers do not
have probabilistic meaning about $\theta$, but they are `confidence intervals'
at such and such `confidence level', because `$\theta$ is 
a constant of unknown value', etc\ldots Good luck!]} 
The situation is simpler and clearer
in the Bayesian approach, in which the above rules $a$ and $b$ do hold too, 
but only as approximations under well defined conditions.
In case the underlying conditions fail 
we know immediately what to do:
\begin{itemize}
\item
restart from Eq.~(\ref{eq:final_gen}) or  from Eq.~(\ref{eq:final_chi2}),
depending on the other underlying hypotheses;
\item
go even one step before Eq.~(\ref{eq:final_gen}), 
namely to the most general Eq.~(\ref{eq:Bayes}), 
if priors matter
(e.g. physical constraints, sensible previous knowledge, etc.).
\end{itemize}
For example, if the $\chi^2$ description of the data was a 
good approximation,
then $f(\theta)\propto e^{-\chi^2/2}$, 
properly normalized, is the solution to the 
problem.\footnote{To be precise, this approximation
is valid if the parameters appear only in the argument of the exponent.
In practice this means that the fitted parameters must not appear in the 
covariance matrix on which the $\chi^2$ depends.
As a simple example in which
this approximation do not hold is that of a linear fit in which 
also the standard deviation $\sigma$ describing the errors along the 
ordinate. The joint inference about line coefficients $m$ and $c$ 
and $\sigma$, having observed $n$ points, 
 is achieved by $f(m,c,\sigma)\propto \sigma^{-n}\,e^{-\chi^2/2}$
(see Sec. 8.2 of Ref.~\cite{BR}). 
}
\begin{figure}
\begin{center}
\begin{tabular}{cc}
 & \\
\epsfig{file=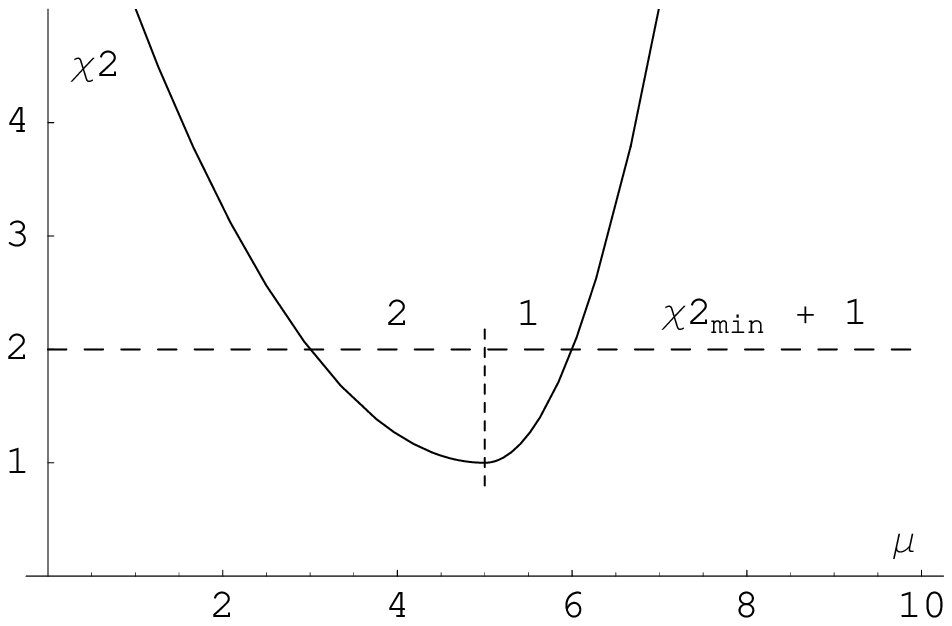,width=0.47\linewidth,clip=} &
\epsfig{file=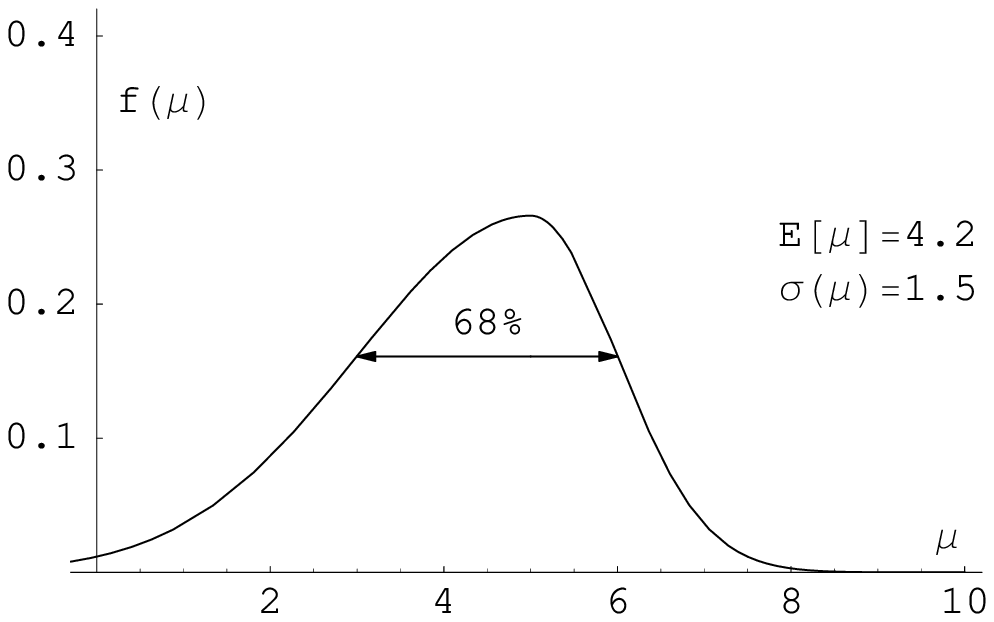,width=0.47\linewidth,clip=} 
\end{tabular}
\end{center}
\caption{\small Example (Ref.~\cite{BR}) of asymmetric $\chi^2$ curve
(left plot)
with a $\chi^2$ minimum at $\mu=5$ ($\mu$ stands for the value
of a generic physics quantity). 
The result based on the $\chi^2_{min}+1$ `prescription' 
is compared (plot on the right side) 
with the p.d.f. based 
on a uniform prior, i.e. 
$f(\mu\,|\,\mbox{data})\propto \exp[-\chi^2/2]$.}  
\label{fig:asymmetric_chi2}
\end{figure}
A non parabolic, asymmetric $\chi^2$ produces an asymmetric 
$f(\theta)$ (see Fig.~\ref{fig:asymmetric_chi2}), 
the mode of which corresponds, indeed, to what obtained
minimizing $\chi^2$, but expected value and standard deviation
differ from what is obtained by the `standard rule'. 
In particular, expected value and variance must be 
evaluated from their definitions:
\begin{eqnarray}
\mbox{E}[\theta] &=& 
 \int\!\theta\,f(\theta\,|\,\mbox{data})\,\mbox{d}\theta 
\label{eq:theta_E} \\
\sigma^2_\theta  &=& 
 \int\!(\theta-\mbox{E}[\theta])^2
       \,f(\theta\,|\,\mbox{data})\,\mbox{d}\theta\,.
\label{eq:theta_sigma}
\end{eqnarray}
Other examples of asymmetric $\chi^2$ curves, including the case
with more than one minimum, are shown in Chapter 12 of Ref.~\cite{BR},
and compared with the
results coming from frequentist prescriptions
(but, indeed, there is not a general accepted rule to
get frequentistic results -- whatever they mean --
when the $\chi^2$ shape gets complicated).

Unfortunately, 
it is not easy to translate numbers obtained by {\it ad hoc}
rules into probabilistic results, because the dependence
on the actual shape of the $\chi^2$ or $\varphi$ curve can be 
not trivial. 
 Anyhow, 
some {\it rules of thumb} can be given in next-to-simple situations
where the  $\chi^2$ or $\varphi$ has only one minimum and 
the  $\chi^2$ or $\varphi$ curve looks like a `skewed parabola', 
like in Fig.~\ref{fig:asymmetric_chi2}:
\begin{figure}
\begin{center}
\epsfig{file=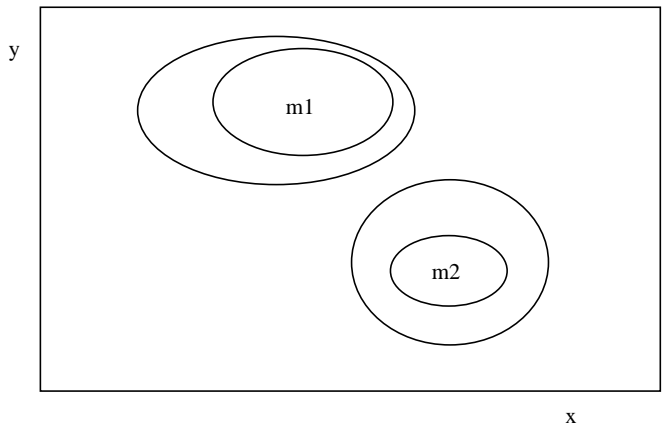,width=0.7\linewidth,clip=}
\end{center}
\caption{\small Example of two-dimensional multi-spots 
``68\% CL'' and ``95\% CL'' contours obtained slicing
 the $\chi^2$
or the minus log-likelihood curve at some 
{\it magic levels}. What do they mean?}
\label{fig:spots}
\end{figure}
\begin{itemize}
\item
the 68\% `confidence interval' obtained by the $\Delta\chi^2=1$,
or $\Delta\varphi=1/2$ rule still provides a 
68\% probability interval for 
$\theta$.
\item
the standard deviation obtained using Eq.~(\ref{eq:theta_sigma}) 
is approximately equal to the average between the $\Delta_+$
and $\Delta_-$ values obtained by the 
 $\Delta\chi^2=1$,
or $\Delta\varphi=1/2$ rule:
\begin{equation}
\sigma_\theta \approx \frac{\Delta_+ + \Delta_-}{2} \,;
\label{eq:sigma_delta_chi2}
\end{equation}
\item
the expected value is equal to the mode ($\theta_m$, 
coinciding with the maximum likelihood or minimum $\chi^2$ value)
plus a {\it shift}:
\begin{equation}
\mbox{E}[\theta] \approx \theta_m + {\cal O}(\Delta_+ - \Delta_-) \,.
\label{eq:shift_chi2}
\end{equation}
[This latter rule is particularly rough because 
$\mbox{E}[\theta]$ 
is more sensitive than $\sigma_\theta$ on the exact shape of 
 $\chi^2$ or $\varphi$ curve. 
Equation (\ref{eq:shift_chi2}) has to
 be taken only to get an idea of the order
of magnitude of the effect. For example, in the case depicted in 
Fig~\ref{fig:asymmetric_chi2} the shift is 80\% of $(\Delta_+ - \Delta_-)$.]
\end{itemize}

The remarks about misuse of $\Delta\chi^2=1$
and $\Delta\varphi=1/2$ rules
can be extended to cases where several parameters are involved.
I do not want to go into details 
(in the Bayesian approach there is 
nothing deeper than studying $k\,e^{-\chi^2/2}$ or  $k\,e^{-\varphi}$ 
in function of several 
parameters.\footnote{See footnote 7 concerning a possible 
pitfall in the use of $k\,e^{-\chi^2/2}$.
}), but I just want
to get  the reader
worried about the meaning of contour plots of the kind shown 
in Fig.~\ref{fig:spots}.

\subsection{Nonlinear propagation}\label{ss:nonlinear}
Another source of asymmetric uncertainties is 
nonlinear dependence of the output quantity $Y$ 
on some of the input $\mvec X$ 
in a
region a few standard deviations 
around $\mbox{E}(\mvec X)$.
 This problem has been studied with great detail
in Ref.~\cite{ConMirko}, 
also taking into account correlations
on input and output quantities,
and somewhat summarized in 
Ref.~\cite{BR}. 
Let us recall here only the most relevant outcomes,
in the simplest case of only one output quantity $Y$
and neglecting correlations.

\begin{figure}[!t]
\begin{center}
\begin{tabular}{|c|c|} \hline
\multicolumn{2}{|c|}{\ } \\
\multicolumn{2}{|c|}{
\epsfig{file=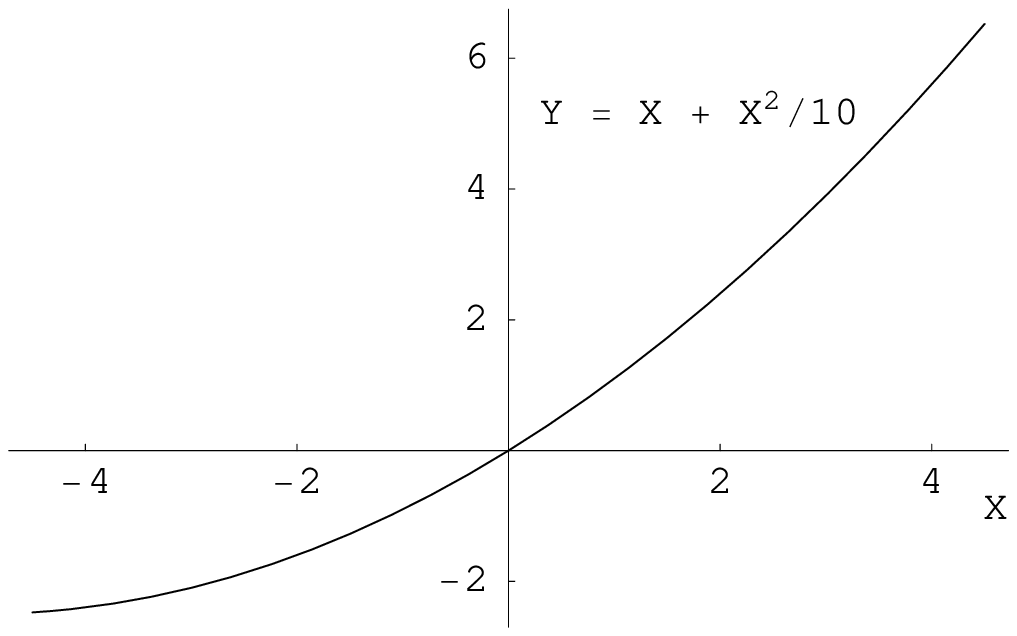,width=0.45\linewidth,clip=}} \\ \hline
& \\
\epsfig{file=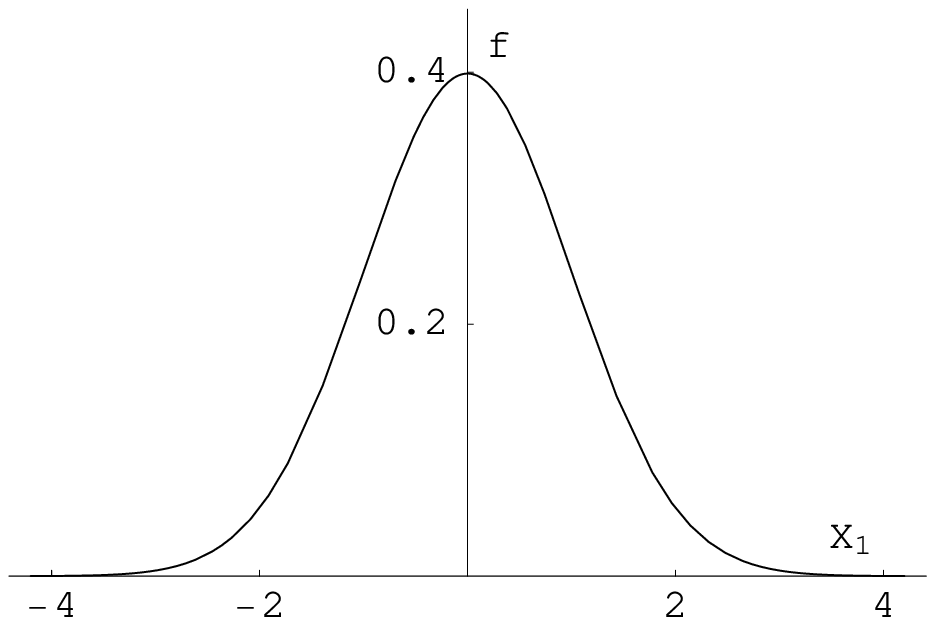,width=0.45\linewidth,clip=} &
\epsfig{file=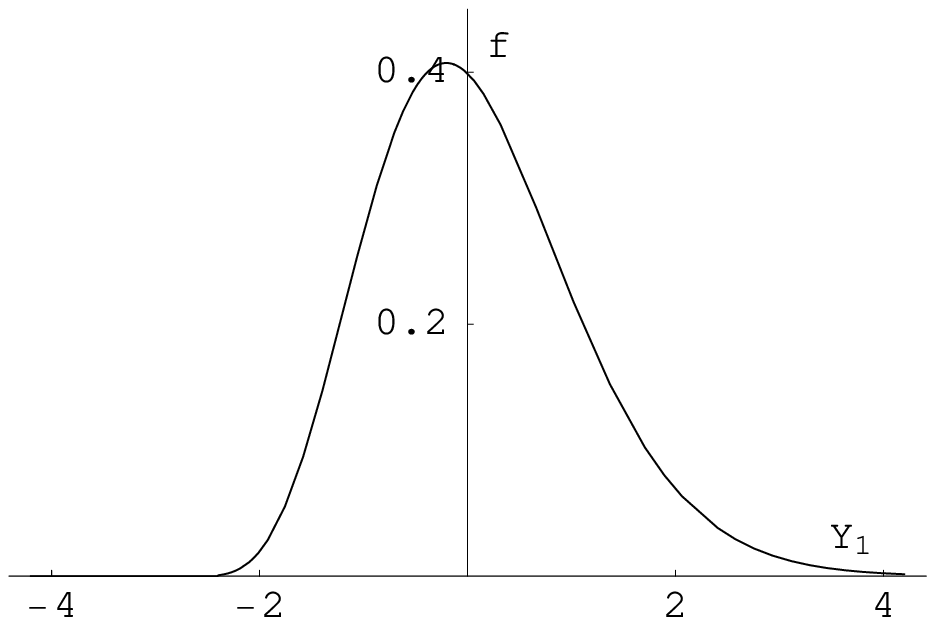,width=0.45\linewidth,clip=}  \\ \hline
\end{tabular}
\end{center}
\caption{\small Propagation of a
Gaussian distribution under a nonlinear transformation. 
$f(Y_i)$ were obtained analytically using  
Eq.(\ref{eq:prop_general}) (part of Fig.~12.2 of Ref.\cite{BR}).}
\label{fig:quad}
\end{figure}
Figure~\ref{fig:quad} shows a non linear dependence between 
$X$ and $Y$ and how a Gaussian distribution has been distorted
by the transformation [$f(y)$ has been 
evaluated analytically using Eq.(\ref{eq:prop_general})].
As a result of the nonlinear transformation, 
mode, mean, median and standard deviation 
are transformed in non trivial ways (in the example of 
Fig.~\ref{fig:quad} mode moves left and expected value right). 
In the general case
the complete calculations should be performed, 
either analytically, or numerically or by Monte Carlo. 
Fortunately, 
as it has been shown in Ref.~\cite{ConMirko}, second order
expansion is often enough to take into account small
deviations from linearity. The resulting formulae
are still compact and depend on location and shape
parameters of the 
initial distributions.

Second order propagation formulae depend on first 
and second derivatives.
In practical cases (especially
as far as the contribution from systematic effects are concerned)
the derivatives are obtained 
numerically\footnote{Note that sometimes people do not get
asymmetric uncertainty, not because the propagation is 
approximately linear,
but because asymmetry is hidden by the standard propagation
formula! Therefore also in this case the approximation might produce
a bias in the result (for example, the second order formula of the
 expected value of the ratio of two quantities is known to 
experts\cite{Mood}). The merit of numerical derivatives
is that at least it shows the  asymmetries. 
}
 as
\begin{eqnarray}
\left.\frac{\partial Y}{\partial X}\right|_{\mbox{\small E}[X]} &\approx& \frac{1}{2}
       \left(\frac{\Delta _+}{\sigma(X)}+
\frac{\Delta _-}{\sigma(X)}\right) =
       \frac{\Delta _+ +\Delta _-}{2\,\sigma(X)}\,, \\
\left.\frac{\partial^2 Y}{\partial X^2}\right|_{\mbox{\small E}[X]}  &\approx &
       \frac{1}{\sigma(X)}\,
       \left(\frac{\Delta _+}{\sigma(X)}-\frac{\Delta _-}{\sigma(X)}\right)
       = \frac{\Delta _+-\Delta _-}{\sigma^2(X)}\,,
\end{eqnarray}
where $\Delta_-$ and $\Delta_+$ 
now stand for the left and right deviations of $Y$ 
when the {\it input variable $X$ varies by one standard deviation}
 around $\mbox{E}[X]$. 
Second order propagation formulae
are conveniently given 
in Ref.~\cite{ConMirko} in terms of the $\Delta_\pm$ 
deviations\footnote{In 
terms of analytically calculated derivatives, 
$\delta$ and $\overline{\Delta}$ are given by
\begin{eqnarray}
\delta &=& \frac{1}{2}\left.\frac{\partial^2 Y}
                                {\partial X^2}\right|_{\mbox{\small E}[X]}
           \, \sigma^2(X) \\
\overline{\Delta} &=& \left.\frac{\partial Y}
                           {\partial X}\right|_{\mbox{\small E}[X]}
           \, \sigma(X)\,.
\end{eqnarray}
}. 
For $Y$ that depends only on a single
input $X$ we get:
\begin{eqnarray}
\mbox{E}(Y) &\approx& Y(\mbox{E}[X]) + \delta\,, 
\label{eq:EY_nonlinear} \\
\sigma^2(Y) &\approx& \overline{\Delta}^2 + 
2\,\overline{\Delta}\cdot\delta\cdot S(X)+
                \delta^2\cdot\left[{\cal K}(X)-1\right]\,,
\label{eq:sig_nonlinear}
\end{eqnarray}
where $\delta$ is the semi-difference of the two deviations
and $\overline{\Delta}$ is their average: 
\begin{eqnarray}
\delta &=& \frac{\Delta _{+}-\Delta _{-}}{2} \\
\label{eq:delta_m}
\overline{\Delta}&=&\frac{\Delta _{+}+\Delta _{-}}{2}\,,
\label{eq:Delta_m}
\end{eqnarray}
while ${\cal S}(X)$ and ${\cal K}(X)$ stand for skewness and  kurtosis
of the input variable.\footnote{After 
what we have seen in Sec.~\ref{sec:propagation} 
we should not forget that the input quantities could have non
trivial shapes. Since skewness and kurtosis are related to 3rd and 4th 
moment of the distribution, Eq.~(\ref{eq:sig_nonlinear})
makes use up to the 4th moment and is definitely better that 
the usual propagation formula, that uses only second moments.
In Ref.~\cite{ConMirko} approximated formulae are given also
for skewness and kurtosis of the output variable, 
from which it is possible to reconstruct $f(y)$ taking into account
up to 4-th order moment of the distribution.
} 

For many input quantities we have 
\begin{eqnarray}
\mbox{E}(Y) &\approx& Y(\mbox{E}[X]) + \sum_i\delta_i\,, 
\label{eq:EY_nonlinear_many} \\
\sigma^2(Y) &\approx& \sum_i\sigma^2_{X_i}(Y)\,,
\label{eq:sig_nonlinear_many}
\end{eqnarray}
where $\sigma^2_{X_i}(Y)$ stands for each individual contribution to
Eq.~(\ref{eq:sig_nonlinear}). The expression of the variance 
gets simplified
when all input quantities are Gaussian 
(a Gaussian has skewness equal 0
and kurtosis equal 3):
\begin{eqnarray}
\sigma^2(Y) &\approx& \sum_i\overline{\Delta}^2_i+2\sum_i\,\delta^2_i\,,
\label{eq:sig_nonlinear_many_Gaussian}
\end{eqnarray}
and, as long as $\delta_i^2$ are much smaller that 
$\overline{\Delta}_i^2$, we get the convenient 
{\it approximated formulae}
\begin{eqnarray}
\mbox{E}(Y) &\approx& Y(\mbox{E}[{\mvec X}]) + \sum_i \delta_i\,, 
\label{eq:nl_simple_E} \\
\sigma^2(Y) &\approx& \sum_i \overline{\Delta}^2_i\,\,
\label{eq:nl_simple}
\end{eqnarray}
valid also for other symmetric input p.d.f.'s (the kurtosis is
about 2 to 3 in typical distribution and its exact value
is irrelevant 
if the condition 
$\sum_i\delta_i^2 \ll \sum_i\overline{\Delta}_i^2$ holds). 
The resulting practical rules 
(\ref{eq:nl_simple_E})--(\ref{eq:nl_simple})
 are quite simple:
\begin{itemize}
\item
the {\it expected value} of $Y$ {\it is shifted}
 by the sum of the individual shifts,
each given by half of the semi-difference of the deviations $\Delta_\pm$;
\item
each input quantity contributes (in quadrature) to the 
combined standard uncertainty
with a term which is approximately the average between the 
deviations $\Delta_\pm$.
\end{itemize}
Moreover, if there are many contributions to the uncertainty, 
the final uncertainty
will be symmetric and approximately Gaussian,
thanks to the central limit theorem. 

\subsection{Uncertainty due to systematics}
Finally, and this is often the case that we see in publications, 
asymmetric uncertainty results from systematic effects.
The Bayesian approach offers a natural and clear way
to treat systematics 
 -- and I smile at the many 
attempts\footnote{It has been studied by psychologists how sometimes
our efforts to solve a problem are the analogous with the 
moves along elements of a group structure (in the mathematical sense).
There is no way to reach a solution until we not
break out of this kind of trapping 
psychological or cultural cages.\cite{WWF}
} of `squaring the circle'
 using frequentistic prescriptions\ldots --
simply because probabilistic concepts are 
consistently applied
to all {\it influence quantities} that can have an effect
on the quantity of interest and whose value is not precisely known.
Therefore we can treat them using probabilistic methods.
This was also recognized by  metrologic
organizations\cite{ISO}. 
 
Indeed, there is no need to treat systematic effects
in a special way. They are treated as any of the many 
input quantities $\mvec X$ discussed in Sec.~\ref{ss:nonlinear},
and, in fact,  their asymmetric contributions come 
frequently from their nonlinear influence on 
the quantity of interest. 
The only word of caution, on which I would like to insist,
is to use expected value and standard deviation
for each systematic effect. 
In fact, sometimes the uncertainty about the value
of the influence quantities that contribute to systematics
is intrinsically asymmetric. 

I also would like to comment shortly on results where
either of the $\Delta_\pm$ is negative, 
for example $1.0^{+0.5}_{+0.3}$ (see e.g. Ref.~\cite{HERA}
to have an idea of the variety of signs of $\Delta_\pm$). This means
that that the we are in proximity of a minimum (or a maximum 
if $\Delta_+$ were negative) of the function $Y=Y(X_i)$. 
It can be shown~\cite{ConMirko,BR} that 
Eqs.~(\ref{eq:EY_nonlinear})-(\ref{eq:sig_nonlinear})
hold for this case too.\footnote{In
this special case there should be
no doubt that a shift should be applied 
to the best value, since moving $X_i$ by $\pm \sigma(X_i)$
around its expected value $\mbox{E}[X_i]$
the final quantity $Y$ only moves 
in one side of $Y(\mbox{E}[X_i])$.}

For further details about meaning and treatment of
uncertainties due systematics and their relations
to ISO {\it Type B} uncertainties\cite{ISO}, see 
Refs.~\cite{ConMirko} and \cite{BR}.

\section{Some rules of thumb to unfold probabilistic sensible information
from results published with asymmetric uncertainties}\label{sec:thumb}
Having understood what one should have done 
to obtain expected value and standard deviation
in the situations in which people are used to report
asymmetric uncertainties, we might attempt to recover
those quantities from the published result. 
It is possible to do it exactly only if we 
know the detailed contributions to the uncertainty, 
namely the $\chi^2$ or log-likelihood functions of the
so called `statistical terms' and the pairs 
$\{\Delta_{+_i}, \Delta_{-_i}\}$, together
to the probabilistic model, 
 for each `systematic term'. 
However, these pieces of information are usually 
unavailable. But we can still make some {\it guesses}, 
based on some rough assumptions, lacking other information:
\begin{itemize}
\item
asymmetric uncertainties in the `statistical part' 
are due to asymmetric $\chi^2$ or log-likelihood:
$\rightarrow$
apply  corrections  given by
 Eqs.~(\ref{eq:sigma_delta_chi2})--(\ref{eq:shift_chi2});
\item
asymmetric uncertainties in the `systematic part'
comes from nonlinear propagation: $\rightarrow$
 apply  corrections  given by
Eqs.~(\ref{eq:nl_simple_E})--(\ref{eq:nl_simple}).
\end{itemize}
As a numerical example, imagine 
 we read the 
following result (in arbitrary units): 
\begin{eqnarray}
Y &=& 6.0\,^{+1.0}_{-2.0}\,^{+0.3}_{-0.9}\,,
\label{eq:esempio_bad_Y}
\end{eqnarray}
(that somebody would summary as $6.0\,^{+1.0}_{-2.2}$!). 
The only certainty we have, seeing
two asymmetric uncertainties with the same 
sign of skewness, is that {\it the result is definitively biased}. 
Let us try to make our estimate of the bias  and 
calculate the corrected result (that, not withstanding all uncertainties
about uncertainties, will be  closer to the
`truth' than the published one): 
\begin{enumerate}
\item
the first contribution gives roughly 
[see. Eqs.~(\ref{eq:sigma_delta_chi2})--(\ref{eq:shift_chi2})]:
\begin{eqnarray}
\delta_1 & \approx & -1.0 \\
\sigma_1 & \approx & 1.5 \,;  
\end{eqnarray}
\item
for the second contribution we have
[see. Eqs.~(\ref{eq:delta_m})--(\ref{eq:Delta_m}), 
(\ref{eq:nl_simple_E})--(\ref{eq:nl_simple})]:
\begin{eqnarray}
\delta_2 & \approx & -0.31 \\
\sigma_2 & \approx & 0.62\,.  
\end{eqnarray}
\end{enumerate}
{\it Our} guessed best result would then 
become\footnote{The ISO Guide~\cite{ISO} 
recommends to give the result using the standard
deviation within parenthesis, instead 
of using the $\pm xx$ notation. 
In this example we would have
$Y\approx 4.69\,(1.5)\,(0.62) = 4.69\,(1.62) 
\Rightarrow Y\approx 4.7\,(1.6)$.
Personally, I do not think this is
a very important issue as long as we know what the quantity
$xx$ means. Anyhow, I understand the ISO rational, 
and perhaps the proposed notation could help to make a break with
the `confidence intervals'.}
\begin{eqnarray}
Y &\approx & 4.69 \pm 1.5 \pm 0.62 = 4.69 \pm 1.62 \\
  &\approx & 4.7 \pm 1.6\,.
\end{eqnarray}
(The exceeding number of digits in the intermediate steps are just  
to make numerical comparison with the correct result that
will be given in a while.)

If we had the chance to learn  that 
the result of Eq.~(\ref{eq:esempio_bad_Y}) was due 
to the asymmetric $\chi^2$ fit of 
Fig.~\ref{fig:asymmetric_chi2} plus two systematic
corrections, each described by the triangular distribution
of Fig.~\ref{fig:2triang}, then we could 
calculate  expectation and variance exactly:
\begin{eqnarray}
\mbox{E}(Y) & = & 4.2 + 2\times 0.17 = 4.54 
\label{eq:exact_corr_E}\\
\sigma^2(Y) & = & 1.5^2 +  2\times 0.42^2 = 1.61^2\,,  
\label{eq:exact_corr_sigma}
\end{eqnarray}
i.e. $Y=4.54\pm 1.61$, 
quite different from Eq.~(\ref{eq:esempio_bad_Y})
and close to the result corrected by  
rule of thumb formulae. Indeed, knowing exactly the 
ingredients,
we can evaluate $f(y)$ from 
Eq.(\ref{eq:prop_general})  as
\begin{eqnarray}
f(y) &=& \int \delta(y - x_1 - x_2 - x_3)\, f_1(x_1)\, 
         f_2(x_2)\, f_3(x_3)\, 
         \mbox{d}x_1\, \mbox{d}x_2\,  \mbox{d}x_3\,,     
\end{eqnarray}
although by Monte Carlo. 
\begin{figure}
\begin{center}
\epsfig{file=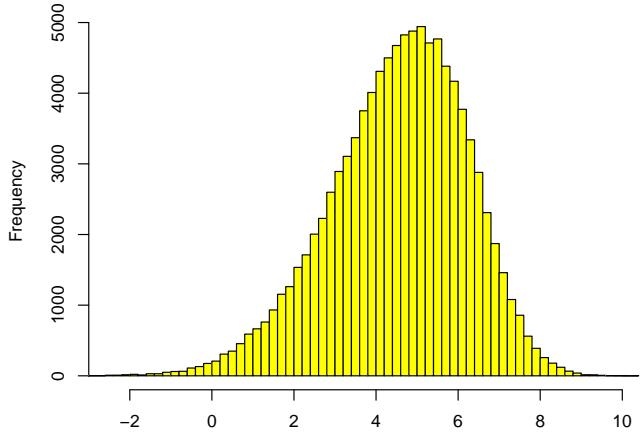,width=0.67\linewidth,clip=}
\end{center}
\caption{\small Monte Carlo estimate of the shape of the
p.d.f. of the sum of three independent variables, 
one described by the p.d.f. of Fig.~\ref{fig:asymmetric_chi2}
and the other two by the 
triangular distribution of  Fig.~\ref{fig:2triang}. 
}
\label{fig:skewed+2tr}
\end{figure}
The result is given in Fig.~\ref{fig:skewed+2tr}, 
from which we can evaluate a mean value of 4.54 and 
a standard deviation of 1.65 in perfect agreement with
the figures given in 
Eqs.~(\ref{eq:exact_corr_E})--(\ref{eq:exact_corr_sigma}).\footnote{The 
slight difference between the standard deviations comes from 
rounding, since $\sigma(\mu)=1.5$ of 
Fig. \ref{fig:asymmetric_chi2} is the rounded value
of 1.54. Replacing 1.5 by 1.54 in Eq.~(\ref{eq:exact_corr_sigma}),
we get exactly the Monte Carlo value of 1.65.} 
As we can see from the figure,
also those who like to think at 'best value'
in term of most probable value have to realize once more
that {\it the most probable value of a sum is not 
necessarily equal to the sum of most probable values of the
addends}
(and analogous statements for all 
combinations of uncertainties\footnote{Discussing this issues with several
persons I have realized, with my great surprise, that this
misconception is deeply rooted and strenuously defended by many colleagues,
even by data analysis experts 
(they constantly reply ``yes, but\ldots'').
This attitude is probably one of the consequences of being 
anchored to what I call un-needed principles 
(namely maximum likelihood, in this case), such   
that even the digits resulting from these
principles are taken with a kind of religious respect
and it seems blasphemous to touch them.}). 
In the distribution of Fig.~\ref{fig:skewed+2tr}, the mode of the distribution
is around 5. [Note that expected value and variance are 
equal to those given
by Eqs.~(\ref{eq:exact_corr_E})--(\ref{eq:exact_corr_sigma}, 
since in the case of a linear combination they can be obtained exactly.]
Other statistical quantities that can
be extracted by the distribution are the median, equal to 4.67,
and some 'quantiles' (values at which the cumulative distribution
reaches a given percent of the maximum --
the median being the 50\% quantile). Interesting quantiles
are the 15.85\%, 25\%, 75\% and 84.15\%, for which the 
Monte Carlo gives the following values of $Y$:
2.88, 3.49, 5.72 and 6.18. From these values we can calculate
the {\it central} 50\% and 68.3\% intervals,\footnote{I give
the central 68.3\% interval with some reluctance, 
because I know by experience that in many minds
the short circuit
$$ 
\mbox{``68\% probability interval'' $\longleftrightarrow$ ``sigma''}
$$
is almost unavoidable (I have known physicists  convinced
-- and who even taught it! --
that the standard deviation only `makes sense for the Gaussian'
and that it was defined via the `68\% rule'). For this
reason, recently I have started to appreciate thinking in terms of 
50\% probability intervals, also  because they force people to reason
in terms of better perceived 
fifty-to-fifty bets. I find these kind of bets very enlighting 
to show why practically all standard ways (including Bayesian ones!)
fail to report upper/lower 
{\it confidence limits} in {\it frontier case situations} characterized
by {\it open likelihoods} (see chapter 12 in Ref.\cite{BR}). I like to ask
``please use your method and give me a 50\% C.L. upper/lower limit'', 
and then, when I have got it,
 ``are you really 50\% confident that the value is below 
that limit and 50\% confident that it is above it?
Would you equally bet on either side of that limit?''. 
And the supporters of `objective' methods
are immediately at loss. (At least those who use Bayesian formulae
realize that there must be some problem with the choice 
of priors.)
}
which are
$[3.49,\ 5.72]$ and $[2.88,\ 6.18]$, respectively.
Again, the information provided by Eq.~(\ref{eq:esempio_bad_Y})
is far from any reasonable way to provide
the uncertainty about $Y$, given the information on each
component.

Besides the lucky case\footnote{In the example here 
we have been lucky because an over-correction
of the first contribution was compensated by an under-correction
of the second contribution. Note also that the hypothesis 
about the nonlinear propagation was not correct, because we had,
instead, a linear propagation of asymmetric p.d.f.'s. Anyhow
the overall shift calculated by the guessed hypothesis
is comparable to that calculable knowing the details of the analysis
(and, in any case, using in subsequent analyses
the roughly corrected result is 
definitely better than sticking to the published `best value').}
of this numerical example (which was not constructed
on purpose, but just recycling  some material from Ref.~\cite{BR}),
it seems reasonable that even results roughly corrected
by rule of thumb formulae are already
better than those published directly with asymmetric
result.\footnote{Note that even if we were told that $Y$ was 
$6.0^{+1.0}_{-2.2}$, without further information, we could
still try to apply some shift to the result, obtaining 
$4.8\pm 1.6$ or $5.4\pm 1.6$ depending on some guesses
about the source of the asymmetry. In any case, either
results are better than $6.0^{+1.0}_{-2.2}$!} 
But the accurate analysis can only
be  done by the authors
who know the details of the individual contribution to the uncertainty.

\section{Conclusions}
Asymmetric uncertainties do exist and there is {\it no
way to  remove them artificially}. If they are 
not properly treated, i.e. using prescriptions 
that do not have a theoretical ground but are 
 more or less rooted in the physics community,
the published result is 
biased.
Instead, if they are properly treated using probability 
theory,
in most cases of interest the final result is 
practically symmetric and approximately Gaussian,
with expected value and standard deviations 
which take into account the several shifts 
due to individual asymmetric contributions.
Note that some of the simplified
methods to make statistical analyses had a 
{\it raison d'\^etre}  many
years ago, when the computation was a serious
limitation. Now it is not any longer a problem  to evaluate,
analytically or numerically, 
 integrals of the 
kind of those appearing e.g. in Eqs.(\ref{eq:prop_general}),  
(\ref{eq:theta_E}) and  (\ref{eq:theta_sigma}). 

In the case the final uncertainty remains asymmetric,
the authors should provide detailed information about the
`shape of the uncertainty', giving also most probable value,
probability intervals, and so on.
But the best estimate of the 
{\it expected value and standard deviation should
be always given} (see also the {\it ISO Guide}~\cite{ISO}). 

To conclude, I would like to leave the final word to my preferred
quotation with whom I like to end seminars and courses
on probability theory applied to the
evaluation and the expression of uncertainty in measurements:
\begin{quote}
{\sl \small
``Although this {\rm Guide} provides a framework for assessing
uncertainty, it cannot substitute for critical
thinking, intellectual honesty, and professional skill. The evaluation
of uncertainty is neither a routine task nor a
purely mathematical one; it depends on detailed knowledge
of the nature of the measurand and of the measurement.
The quality and utility of the uncertainty quoted for the result of a
 measurement therefore ultimately depend on the understanding,
 critical analysis, and integrity of those who contribute to
 the assignment of its value.''}\cite{ISO} 
\end{quote}

\vspace{1.0cm}
\noindent
It is a pleasure to thank {\it Superfaber} 
(Fabrizio Fabbri in {\it hepnames}) for helpful 
discussions on the subject and for his {\it super}vision of the
manuscript.

\end{document}